\documentclass[12pt]{article}
\usepackage{amsmath}

\def\Zop{\bbbz}

\def\bbbz {{\sf Z\!\!Z}}

\def\be{\begin{equation}}
\def\ee{\end{equation}}
\def\ba{\begin{eqnarray}}
\def\ea{\end{eqnarray}}
\def\p{\partial}
\def\ap{\alpha^\prime}
\def\I{{\cal I}}
\def\N{{\bf N}}
\def\R{{\bf R}}
\def\Tr{\mbox{Tr}}

\def\xb{{\bar X}}
\numberwithin{equation}{section} 
\parskip=\medskipamount

\begin{document}

\thispagestyle{empty}
\def\thefootnote{\fnsymbol{footnote}}
\begin{flushright}
  hep-th/0312091 \\
  SPIN-2003/45 \\
  ITP-2003/64
\end{flushright}

\vskip 0.5cm

\begin{center}\LARGE
{\bf Open Spinning Strings}
\end{center}

\vskip 1.0cm

\begin{center}
{\large B. Stefa\'nski, jr.\footnote{E-mail address: {\tt stefansk@phys.uu.nl}}}

\vskip 0.5cm

{\it $^*$ Spinoza Institute, University of Utrecht \\
Postbus 80.195, 3508 TD Utrecht, The Netherlands}
\end{center}

\vskip 1.0cm

\begin{center}
December 2003
\end{center}

\vskip 1.0cm

\begin{abstract}
\noindent We find classical open string solutions in the $AdS_5\times S^5/\Zop_2$ orientifold with angular momenta along the five-sphere. The energy of these solutions has an expansion in integral powers of $\lambda$ with sigma-model corrections suppressed by inverse powers of $J$ - the total angular momentum. This gives a prediction for the exact anomalous dimensions of operators in the large $N$ limit of an ${\cal N}=2$ $Sp(N)$ Super-Yang-Mills theory with matter.
We also find a simple map between open and closed string solutions. This gives a prediction for an all-loop planar relationship between the anomalous dimensions of single-trace and two-quark operators in the dual gauge theory.
\end{abstract}

\vfill

\setcounter{footnote}{0}
\def\thefootnote{\arabic{footnote}}
\newpage

\renewcommand{\theequation}{\thesection.\arabic{equation}}
\section{Introduction}
Recently there has been a great deal of interest in stringy tests of the AdS/CFT correspondence~\cite{adscft}.\footnote{For a review see~\cite{rev}.} This work was initiated by the discovery of the maximally supersymmetric Type IIB plane-wave background~\cite{bfhp}, and its subsequent light-cone quantisation~\cite{ppq}. 
Since the plane wave background can be viewed as a Penrose limit of $AdS_5\times S^5$~\cite{bfhp}, in~\cite{bmn} it was argued that the anomalous dimensions $\Delta$ of a large R-charge $J$ sector of the ${\cal N}=4$ $SU(N)$ SYM theory were related to the energies $E$ of single string states in the plane-wave background.
\be
E=\Delta-J\,.
\ee
 The duality was first proposed~\cite{bmn} and investigated~\cite{ppp} in the planar limit, at all loops for operators of large R-charge $J$. Subsequently, through string interactions in the plane wave background~\cite{ppsft}, the non-planar corrections to this were considered~\cite{ppnp}.\footnote{For a different approach to this see~\cite{divecchia}.} In~\cite{gkp2,FT1,pr,CS} $1/J$ corrections to the BMN duality were considered, which, on the string side, correspond to going back towards the $AdS_5\times S^5$ geometry from the plane-wave background.

\noindent In the original large $N$ expansion in gauge theory~\cite{th},\footnote{For the large $N$ limits of $SO$ and $Sp$ gauge theories relevant to this paper see~\cite{cicuta}.} $1/N^2$ was used as an expansion parameter. An essential feature of the BMN duality was the presence of a large R charge, allowing for $1/J$, as well as $1/N^2$, to be used as a small parameter. In this approach one considers classical spinning  string solutions in the $SO(2,4)\times SO(6)$ sigma model, which carry a large amount of angular momentum or spin in some directions~\cite{gkp2,FT1,abp}. The energy of such a configuration is then determined as an expansion in angular momenta and $\alpha^\prime=1/\sqrt{\lambda}$, and can be compared with the planar anomalous dimensions of gauge theory operators which transform in the corresponding representation of the superconformal group.

\noindent Recent progress~\cite{FT2,FT3,FT4,AFRT,ART} has been made by studying closed string solutions which carry angular momentum in the $S^5$ directions. It has been argued that the energy of such solutions has a double expansion in $\lambda/J^2$ and $1/J$
\begin{equation}
E(J)=J\left[1+\sum_{k=1}^\infty\left(\frac{\lambda}{J^2}\right)^2\left(c_k+\sum_{n=1}^\infty\frac{d_{nk}}{J^n}\right)
\right]\,,\label{eexp}
\end{equation}
where $J$ is the total angular momentum of the solution on $S^5$. This expansion has the beautiful feature that
the classical energy, expressed as a function of $J$ and string tension $\sqrt{\lambda}$, arranges itself into a regular expansion in $\lambda/J^2$ with quantum sigma-model corrections suppressed by $1/J$. This allows for exact comparisons with gauge theory. Following the work on spin chains, the Bethe ansatz, and ${\cal N}=4$ SYM~\cite{mz,gt} extensive comparisons with gauge theory have been made~\cite{test,FT4,AFRT,ART}, and impressive agreement has been found.\footnote{It should also be noted that integrable structures have been discussed in non-supersymmetric gauge theories as well~\cite{nonsusyint}.} Further, in~\cite{AS}, a beautiful relation was established between an infinite set of local charges on both sides of the duality (for work on non-local charges see~\cite{nonloc}.). It would be interesting to understand the general criteria for which classical sigma model solutions exhibit these nice properties.~\footnote{For example, based on present evidence, these properties seem to be connected to the presence of a large R charge. Solutions with large spin, but small angular momenta, seem not to possess these nice properties.}

\noindent In this paper we find classical spinning open string solutions whose energies do have an expansion of the form~(\ref{eexp}). In particular we consider the $AdS_5\times S^5/\Zop_2$ orientifold~\cite{ororf}, which can be thought of as the near-horizon geometry of a tadpole-canceling O7-plane D7-brane system in the presence of a large number, $2N$, of D3-branes. The metric on $AdS_5\times S^5$ is given by
\ba
ds^2&=&-\cosh^2\rho dt^2+d\rho^2+\sinh^2\rho(d\theta^2+\sin^2\theta d\phi+\cos^2\theta d\varphi^2) \nonumber \\
&&+d\gamma^2+\cos^2\gamma d\varphi_1^2+\sin^2\gamma(d\psi^2+\cos^2\psi d\varphi_2^2+\sin^2\psi d\varphi_3^2)
\,.
\ea
The orientifold action is generated by $g\equiv\Omega\I_Y(-1)^{F_l}$, where $\Omega$ acts as
\be
\Omega(\sigma)=\sigma_{\mbox{\scriptsize max}}-\sigma\,,\label{omegac}
\ee
with $\sigma_{\mbox{\scriptsize max}}=\pi,2\pi$ for open and closed strings, respectively;
$(-1)^{F_l}$ (needed for the orientifold action to square to one on fermions) acts trivially on the bosonic solutions considered here and $\I_Y$ acts trivially on all coordinates of $AdS_5\times S^5$ apart from $\varphi_3$ on which it acts as
\be
I_Y(\varphi_3)=\varphi_3+\pi\,.
\ee
The fixed point is then at $\psi=0$. In terms of global $X,\,Y,\,Z$ coordinates on the sphere
\be
Z=\cos\gamma e^{i\varphi_1}\,,\qquad
X=\sin\gamma\cos\psi e^{i\varphi_2}\,,\qquad
Y=\sin\gamma\sin\psi e^{i\varphi_3}\,,\label{sphcoords}
\ee
$\I_Y$ acts trivially on $X$ and $Z$ and on $Y$ acts as
\be
\I_Y(Y)=-Y\,,\label{Iy}
\ee
and the fixed point is located at $Y=0$. As a result the O7-planes and D7-branes stretch over the $AdS_5$ space as well as $X$ and $Z$. In other words the open strings will have Neumann boundary conditions along
$X$ and $Z$ with $Y$ Dirichlet. The $SO(6)$ R-symmetry is broken to $SO(4)\times SO(2)$ by the orientifold projection, and half of supersymmetry is broken by the O7-plane and D7-branes; further, the $SU(2N)$ gauge group on the world-volume of the D3-branes is broken to $Sp(N)$. In the large $N$ limit the surviving states are the D3-D3 and D3-D7 open strings. The former give rise to an ${\cal N}=2$ theory with a hypermultiplet in the antisymmetric representation of $Sp(N)$, while the latter give rise to four hypermultiplets in the fundamental of $Sp(N)$. The $SO(8)$ gauge group on the D7-branes' world-volume becomes a global symmetry of the gauge theory. At the origin of moduli space the D3-brane theory has been shown to be conformal~\cite{spth}. The BMN sector of this theory was investigated in~\cite{pporf,gmp}.

\noindent This paper is organised as follows. In section~\ref{sec2} we briefly review classical spinning open and closed string solutions in orientifolds of flat space.
In section~\ref{sec3} we construct orientifold invariant open and closed string solutions of the circular and folded type~\cite{FT2,FT3,FT4}. In section~\ref{sec6} we consider general Neumann model type solutions~\cite{AFRT} in this geometry. We find a simple relationship between orientifold-invariant closed and open string solutions~(\ref{opentoclosed}) and their energies~(\ref{eoec}). As a result, for all solutions found, the energy has the desired expansion~(\ref{eexp}). This provides a prediction for the anomalous dimensions of large R-charge operators on the gauge theory side, as well as prediction for a relationship between the planar anomalous dimensions of single-trace and two-quark operators. We briefly discuss the dual gauge theory in
section~\ref{sec7} and we conclude in section~\ref{sec8}. An appendix is included in which we study the stability of open and closed string circuar type solutions in the orientifold model.
\section{Flat spacetime classical rotating strings}\label{sec2}
In this section we briefly review classical rotating open and closed string solutions in flat space, and analyse
their behaviour under orientifold actions. To simplify the notation we restrict to $\R^{1,4}$.
The string coordinates satisfy
\be
(\p_\tau^2-\p_\sigma^2)X_M(\tau,\sigma)=0\,,\label{eom}
\ee
and the constraints
\be
{\dot X}_M{\dot X}_M+X^\prime_MX^\prime_M =0\,,\qquad
{\dot X}_MX^\prime_M =0\,.\label{constr}
\ee
The closed string periodicity condition is
\be
X(\tau,\sigma+2\pi)=X(\tau,\sigma+2\pi)\,.\label{period}
\ee
The solution describing a classical string rotating in two planes is
\be
X_0=\kappa\tau\,,\qquad
X=X_1+iX_2=a_1\cos(n_1\sigma)e^{in_1\tau}\,,\qquad 
Y=X_3+iX_4=a_2\sin[n_2(\sigma+\sigma_0)]e^{in_2\tau}\,,
\ee
for $n_1,\,n_2\in\N$ and $a_1,\,a_2,\,\sigma_0\in\R$ and
\be
\kappa^2=n_1^2a_1^2+n_2^2a_2^2\,.
\ee
The energy and spins are
\ba
E&=&\frac{1}{2\pi\alpha^\prime}\int_0^{2\pi}d\sigma{\dot X}_0=\frac{\kappa}{\alpha^\prime}\,,\label{energy}\\
J_1&=&\frac{i}{4\pi\alpha^\prime}\int_0^{2\pi}d\sigma(X{\dot{\bar X}}-{\bar X}{\dot X})=\frac{n_1a_1^2}{2\alpha^\prime}\,,\label{angmom}\\
J_2&=&\frac{i}{4\pi\alpha^\prime}\int_0^{2\pi}d\sigma(Y{\dot{\bar Y}}-{\bar Y}{\dot Y})=\frac{n_2a_2^2}{2\alpha^\prime}\,.
\ea
In other words
\be
E=\sqrt{\frac{2}{\ap}(n_1J_1+n_2J_2)}\,.
\ee

\noindent We will be interested in open strings which satisfy Neumann or Dirichlet boundary conditions at their endpoints
\ba
\mbox{N:}\qquad\p_\sigma X_M(\tau,\sigma=0,\pi)&=&0\,,\label{N}\\
\mbox{D:}\!\!\!\!\!\qquad\qquad X_M(\tau,\sigma=0,\pi)&=&0\,,\label{D}
\ea
as well as the equation of motion~(\ref{eom}) and the constraints~(\ref{constr}). 
In particular, for $X_0$ and $X$ Neumann and $Y$ Dirichlet, the solution is
\be
X_0=\kappa\tau\,,\qquad
X=X_1+iX_2=a_1\cos(n_1\sigma)e^{in_1\tau}\,,\qquad 
Y=X_3+iX_4=a_2\sin(n_2\sigma)e^{in_2\tau}\,.
\ee
with
\be
\kappa^2=n_1^2a_1^2+n_2^2a_2^2\,.
\ee
In fact the full open string solution is given by $X_M\otimes\beta$, where $\beta$ is a Chan-Patton matrix.
The energy and spins are given by integrals as in the closed string case above, with the integration range now being $0\le\sigma\le\pi$, from which we find that
\be
E=\sqrt{\frac{1}{\ap}(n_1J_1+n_2J_2)}\,.
\ee

\noindent Finally, the orientifold will act by $g\equiv\Omega\I_{34}(-1)^{F_l}$, as well as on the Chan-Patton matrix by
\be
g(\beta)=\gamma_g\beta^T\gamma_g^{-1}\,,\label{ocp}
\ee
where $\gamma_g$ is a matrix satisfying the usual consistency conditions~\cite{ocons}.
It is an easy check to see that the closed string solution is $g$-invariant provided that
$\sigma_0=0,\pi,\dots$ (we might as well set $\sigma_0=0$ by re-defining $a_2$).
The open string solution $X_M\otimes\beta$ is $g$ invariant for $n_1,\,n_2$ even,
and $\beta=\gamma_g\beta^T\gamma_g^{-1}$. A $g$-invariant solution with $n_1,\,n_2$ odd is obtained by
swapping the $X$ and $Y$ coordinates.
\section{Two Spin String Solutions on $S^5$}\label{sec3}
In this section we extend the two spin circular~\cite{FT2,FT3} and folded solutions~\cite{FT4} to open and closed string solutions in our orientifold model. 
\subsection{Circular solutions}\label{sec31}
The major difference to flat space classical string solutions is that now the co-ordinates satisfy the constraint
\be
X_AX_A=|X|^2+|Y|^2+|Z|^2=1\,,\label{sphconstr}
\ee
where
\be
Z=X_1+iX_2\,,\qquad X=X_3+iX_4\,,\qquad Y=X_5+iX_6\,,
\ee
This constraint is enforced by a Lagrange multiplier ${\tilde\Lambda}$ with the Lagrangian
\be
L=\p_aX_A\p^aX_A+{\tilde\Lambda}(X_AX_A-1)\,.
\ee
The equations of motion then are
\be
-\p^2X_A+{\tilde\Lambda} X_A=0\,,\qquad X_AX_A=1\,,\qquad {\tilde\Lambda}=-\p_aX_A\p^aX_A\label{spheoms}
\ee
as well as the contraints~(\ref{constr}), with the index $M$ now running over time and the six fields on $S^5$. Finally,
the closed string solution is periodic while the open string one satisfies Neumann boundary conditions in the $X,Y$ directions an Dirichlet boundary conditions in the $Z$ direction. 

\noindent The closed string solution with a constant Lagrange multiplier ${\tilde\Lambda}=\nu^2$ is~\cite{FT2}
\be
X_0=\kappa\tau\,,\qquad X=\sin\gamma_0\cos(n\sigma) e^{iw\tau}\,,\qquad Y=\sin\gamma_0\sin(n\sigma) e^{iw\tau}\,,\qquad
Z=\cos\gamma_0e^{i\nu\tau}\,,\label{clsoln}
\ee
where $n$ is an integer and $\nu,w,\kappa$ and $\gamma_0$ are real numbers which satisfy
\be
w^2=n^2+\nu^2\,,\qquad \sin^2\gamma_0=\frac{1}{2n_1^2}(\kappa^2-\nu^2)\,.
\ee
From equations~(\ref{energy}),~(\ref{angmom}) it follows that the above solution has 
\be
E=\sqrt{\lambda}\kappa\,,\qquad
J_Z=\sqrt{\lambda}\nu[1-\frac{\kappa^2-\nu^2}{2n^2}]\,,\qquad
J_X=J_Y=\frac{\sqrt{\lambda}}{4}\sqrt{1+(\nu/n)^2}(\kappa^2-\nu^2)\,.
\ee

\noindent The closed string solution above can be interpreted as an open string solution (with $\sigma\in[0,\pi]$ now) for a string with Neumann boundary conditions in the $X$ and $Z$ directions and Dirichlet boundary conditions in $Y$.
The full classical rotating open string solution should be tensored with $\beta$, the Chan-Patton matrix.

\noindent It is easy to see that the closed string solution~(\ref{clsoln}) is invariant under the orientifold generator $g\equiv{\cal I}_Y\Omega(-1)^{F_l}$. Note, however, that circular closed string solutions with a different orientation 
(obtained by permuting $X,Y$ and $Z$) are not $g$-invariant. On the open string solution $g$ acts as
\be
g(X_0)=X_0\,,\qquad g(X)=(-1)^{n_1}X\,,\qquad
g(Y)=(-1)^{n_2}Y\,,\qquad g(Z)=Z\,,\label{gopen}
\ee
As in flat space, the open string solution carries Chan-Patton indices $\beta$, on which $g$ acts as in equation~(\ref{ocp}). These Chan-Patton factors give rise to an $SO(8)$ R-symmetry group on the gauge theory side. We see immediately that the solution is $g$-invariant for $n$ even and $\beta=\gamma_g\beta^T\gamma_g^{-1}$. Such open string solutions have the following energy and angular momenta
\be
E=\frac{\sqrt{\lambda}\kappa}{2}\,,\qquad
J_Z=\frac{\nu\sqrt{\lambda}}{2}[1-\frac{\kappa^2-\nu^2}{2n^2}]\,,\qquad
J_X=J_Y=\frac{\sqrt{\lambda}}{8}\sqrt{1+(\nu/n)^2}(\kappa^2-\nu^2)\,,
\ee
with $n$ even. Swapping the $X$ and $Y$ coordinates gives a $g$-invariant solution for $n$ odd.

\noindent In~\cite{FT3} small fluctuations around closed string circular solutions were analysed, in order to study the stability of the circular solutions. It was found that only for certain ranges of the parameters were the solutions stable. 
One might hope that the orientifold projection will stabalise some of the solutions which were unstable in~\cite{FT3}. In Appendix~\ref{sec4} we carry out the relevant analysis for both open and closed strings. Unfortunately, we find that while the orientiation projection removes some of the unstable modes, it does not remove all of them. As a result the regions of stability of the circular solutions analysed in this section are not improved compared to the unorbifolded theory.
\subsection{Folded-type solutions}\label{sec32}
In this section we review the folded rotating string solution of~\cite{FT4}, and extend it to the case of an open string.
The classical folded string rotating in two planes is given by
\be
t=\kappa\tau\,,\qquad\rho=0\,,\qquad\gamma=\pi/2\,,\qquad\varphi_1=0\,,\qquad\varphi_2=w_1\tau\,,\qquad\varphi_3=w_2\tau\,,\qquad\psi=\psi(\sigma)\,.\label{folded}
\ee
where
\be
\psi^{\prime\prime}+\frac{1}{2}w_{21}^2\sin 2\psi=0\,.
\ee
Integrating once we get
\be
\psi^{\prime 2}(\sigma)=w_{21}^2(\sin^2\psi_0-\sin^2\psi(\sigma))+k^2\,,\label{2spinode}
\ee
with $w_{21}^2\equiv w_2^2-w_1^2$ is taken to be positive, and
\be
-\psi_0\le\psi(\sigma)\le\psi_0\,.\label{foldstab}
\ee
The folded string solutions have $k=0$, while the circular string solutions have $w_{21}=0$. Solutions with $k$ and $w_{21}$ both non-zero are expected to be unstable~\cite{gkp2,FT4}.
This solution has energy $\sqrt{\lambda}\kappa$ which satisfies due to the conformal gauge constraint
\be
E=\sqrt{\lambda}\kappa=\sqrt{\lambda}\sqrt{w_2^2\sin^2\psi_0+w_1^2\cos^2\psi_0}\,,\label{folde}
\ee
and carries angular momenta
\be
{\cal J}_X=w_1\int_0^{2\pi}\frac{d\sigma}{2\pi}\cos^2\psi\,,\qquad
{\cal J}_Y=w_2\int_0^{2\pi}\frac{d\sigma}{2\pi}\sin^2\psi\,.\label{j2}
\ee
The closed string solution is periodic
\be
\psi(\sigma+2\pi)=\psi(\sigma)\,,
\ee
and so for the $n$-folded string 
\be
2\pi=\int_0^{2\pi}d\sigma=4n\int_0^{\psi_0}\frac{d\psi}{w_{21}\sqrt{\sin^2\psi_0-\sin^2\psi}}\,.\label{foldperiod}
\ee
This determines $\psi_0$ in terms of $w_1$ and $w_2$, which are in turn functions of $J_1,J_2$ through 
equation~(\ref{j2}); the energy as a function of the angular momenta follows from equation~(\ref{folde}).

Consider next open string solutions with two Neumann and one Dirichlet directions. The solution~(\ref{folded}) satisfies 
\be
\p_\sigma\psi(\sigma=0,\pi)=0\,,
\ee
and since $\gamma=\pi/2$, we see from equation~(\ref{sphcoords}) that we may interpret~(\ref{folded}) as an open string solution with Neumann boundary conditions in $X$ and $Y$ directions and Dirichlet boundary conditions in the $Z$ direction.\footnote{To make it consistent with the orientifold action, we should redefine $g$ to act by $\Omega(-1)^{F_l}{\cal I}_Z$ rather than by $\Omega(-1)^{F_l}{\cal I}_Y$.} Such an open string has energy $\sqrt{\lambda}\kappa/2$ and carries angular momenta ${\cal J}_1$ and ${\cal J}_2$ given by equation~(\ref{j2}) with the integration range changed to $0\le\sigma\le\pi$.
The open string solution also satisfies equation~(\ref{foldperiod}). Since $Z\equiv 0$, we may also regard it as satisfying Neumann boundary conditions. One might wonder then if a folded-type solution ({\em i.e.} with $k=0$) can be found for $X$ and $Z$ satisfying Neumann and $Y$ Dirichlet boundary conditions. It is not difficult to check that there are no such solutions.

\noindent It is easy to see that the closed string folded solutions considered above are invariant under the orientifold action
$g\equiv{\cal I}_Z\Omega(-1)^{F_l}$, in other words for folded closed strings which do not extend in the Dirichlet direction. Similarily the open string solution is also invariant under the orientifold action if it too does not extend in the Dirichlet direction and
its end-points are at the same point on $S^5$ - in other words for $n$ even.
\section{The Three Spin Solutions}\label{sec6}
\noindent In the previous sections we considered the simplest circular and folded solutions. In this section we now turn to the general class of solutions studied in~\cite{AFRT}. These are of the form 
\be
X_1+iX_2=x_1(\sigma)e^{iw_1\tau}\,,\qquad
X_3+iX_4=x_2(\sigma)e^{iw_2\tau}\,,\qquad
X_5+iX_6=x_3(\sigma)e^{iw_3\tau}\,,
\ee
where
\be
\sum_{i=1}^3x_i^2=1\,.\label{sphere}
\ee
They carry charges $J_i$ under the Cartan subalgebra of $SO(6)$
\be
J_i=\sqrt{\lambda}w_i\int_0^{2\pi}\frac{d\sigma}{2\pi}x^2_i(\sigma)\,,\qquad i=1,\,2,\,3,
\ee
while the energy is~\footnote{These definitions are for closed string charges; the open string charges are the same but with the integration range now being $0\le\sigma\le\pi$; the open string energy is $\sqrt{\lambda}\kappa/2$.}
\be
E=\sqrt{\lambda}\kappa\,,
\ee
with the Virasoro constraint
\be
\kappa^2={\dot X}_M{\dot X}_M+X_M^\prime X_M^\prime\,,
\ee
which makes the energy a function of the three spins. The Lagrangian for the three fields $x_i(\sigma)$, together with the 
Lagrange multiplier enforcing~(\ref{sphere}) is
\be
L=\frac{1}{2}\sum_{i=1}^3(x^{\prime 2}_i-w_i^2x^i_2)+\frac{1}{2}\Lambda(\sum_{i=1}^3x_i^2-1)\,.\label{xlag}
\ee
The closed string solution satisfies
\be
x_i(\sigma)=x_i(\sigma+2\pi)\,,\qquad i=1,2,3\label{closedbc}
\ee
while the open string solution will satisfy
\be
\p_\sigma x_a(\sigma=0,\pi)=0\,,\qquad a=1,\,2\qquad \qquad x_3(\sigma=0,\pi)=0\,.\label{openbc}
\ee
In~\cite{AFRT} it was found that elipsoidal coordinates are useful in describing the closed string system. These coordinates are 
the two real roots $\zeta_1$ and $\zeta_2$ which solve the quadratic equation
\be
\frac{x_1^2}{\zeta-w_1^2}+\frac{x_2^2}{\zeta-w_2^2}+\frac{x_3^2}{\zeta-w_3^2}=0\,.
\ee
The Lagrangian for $\zeta_1,\,\zeta_2$ is
\be
L=\frac{1}{2}g_{ab}(\zeta)\zeta_a^\prime\zeta_b^\prime -U(\zeta)\,,\label{zetalag}
\ee
where the metric is diagonal
\be
g_{11}=\frac{\zeta_2-\zeta_1}{4(\zeta_1-w_1^2)(\zeta_1-w_2^2)(\zeta_1-w_3^2)}\,,\qquad
g_{22}=\frac{\zeta_1-\zeta_2}{4(\zeta_2-w_1^2)(\zeta_2-w_2^2)(\zeta_2-w_3^2)}\,,
\ee
and the equations of motion are
\be
\left(\frac{d\zeta_i}{d\sigma}\right)^2=-4\frac{P(\zeta_i)}{(\zeta_2-\zeta_1)^2}\,,\qquad\mbox{where }
P(\zeta)=(\zeta-w_1^2)(\zeta-w_2^2)(\zeta-w_3^2)(\zeta-b_1)(\zeta-b_2)\,.\label{eoms}
\ee

\noindent The closed string solution will then obviously satisfy
\be
\zeta_1(\sigma+2\pi)=\zeta_1(\sigma)\,,\qquad \zeta_2(\sigma+2\pi)=\zeta_2(\sigma)\,.
\ee
\noindent We may ask what boundary conditions are consistent with the vanishing of the boundary variation of the Lagrangian~(\ref{zetalag}). Such terms come from the integration by parts of the variation of the $(\zeta^\prime)^2$ terms 
and they vanish for
\be
\zeta_2(\sigma=0)-\zeta_1(\sigma=0)=0\,,\label{weirdbczeta}
\ee
or $\zeta_1$ and $\zeta_2$ satisfying the usual Neumann or Dirichlet boundary conditions
\be
\zeta^\prime_i(\sigma=0)=0\,,\qquad \zeta_i(\sigma=0)=cst\,,\qquad i=1,2\,.
\ee
The boundary conditions~(\ref{weirdbczeta}) do not seem compatible with the equations of motion~(\ref{eoms}) and so we will not consider them here.

\noindent For $x_i\ge 0$ and $\zeta$ in the range
\be
w_1^2\le\zeta_1\le w_2^2\le\zeta_2\le w_3^2\,,
\ee
we have
\be
x_i=\sqrt{\frac{(\zeta_1-w_i^2)(\zeta_2-w_i^2)}{w_{ji}^2w_{ki}^2}}\,,\qquad w^2_{ij}\equiv w_i^2-w_j^2\,,
\ee
for $i=1,\,2,\,3$ and $j,\,k$ such that $\epsilon_{ijk}\neq 0$.
We are interested in $x_3$ satisfying Dirichlet boundary conditions. From the above it is easy to see that this is equivalent to 
imposing (in our range of $\zeta_i$) at the boundary
\be
\zeta_2(\sigma=0,\pi)=w_3^2\,.
\ee
Next, with the above boundary condition on $\zeta_2$, requiring that $x_1$ and $x_2$ satisfy Neumann boundary conditions is 
equivalent to requiring
\be
\zeta_1=b_1\mbox{ or }b_2\,,
\ee
at the endpoint. Either of these boundary conditions ensures that $0=\p_\sigma x_1=\p_\sigma x_2=x_3$ with $\p_\sigma x_3\neq 0$ on the boundary. The open string starts on the equator of the $S^3$ parametrised by $\zeta_i$ and moves off north or south. It will at some point reach maximum longitude and then move back down towards the equator/brane. The above applies at both endpoints $\sigma=0,\,\pi$ of the string.

There are of course many types of open string solutions with the above boundary conditions. A solution is characterised by the range of $\zeta_1$ and $\zeta_2$ and by the type and number of extremal points along it. An extremal point is any point $\sigma_k$ for which $x_i^\prime(\sigma_k)=0$ for some $i=1,2,3$. It is easy to see that there are three types of extremal points
\begin{itemize}
\item {\bf F}: A {\em fold} point has all $x_i^\prime=0$; here $\zeta_1,\zeta_2=b_1,b_2$;
\item {\bf B$_i$}: A {\em bend} point has $x_i^\prime\neq 0$ and $x_j^\prime=0$ for $j\neq i$; here one of the $\zeta's$ equals $w_i^2$ while the other $b_a$;
\item {\bf A}: An {\em arch} point has $x_i^\prime=0$ and $x_j^\prime\neq 0$ for $j\neq i$; here $\zeta_1,\zeta_2=w_j^2,w_k^2$, for $j,k\neq i$.
\end{itemize}
Given our choice $\zeta_1\le\zeta_2$ there are four possible ranges for the solutions
\begin{itemize}
\item {\bf I}: $b_1\le\zeta_1\le w_2^2$, $b_2\le\zeta_2\le w_3^2$;
\item {\bf II}: $b_1\le\zeta_1\le b_2$, $w_2^2\le\zeta_2\le w_3^2$;
\item {\bf III}: $w_1^2\le\zeta_1\le w_2^2$, $b_2\le\zeta_2\le w_3^2$;
\item {\bf IV}: $w_1^2\le\zeta_1\le b_2$, $w_2^2\le\zeta_2\le w_3^2$.
\end{itemize}
The extremal points allowed in each range are
\begin{equation}\mbox{{\bf I}: F, B$_2$, B$_3$, A$_1$,}\qquad
\mbox{{\bf II}: B$_2$, B$_3$,}\qquad
\mbox{{\bf III}: B$_1$, B$_2$, A$_1$, A$_2$,}\qquad
\mbox{{\bf IV}: B$_2$, B$_3$, A$_2$, A$_3$,}
\label{allowedextrema}
\end{equation}
From the boundary conditions~(\ref{openbc}), we see that the endpoints of the open string are B$_3$ extremal points. As a result the range III is of no interest to us. A general open string solution will then start and end with the extremal point $B_3$, and have various intermediate extremal points allowed by its $\zeta$ range ({\em c.f.} equation~(\ref{allowedextrema})). As in~\cite{AFRT}, one can consider limiting cases in which one of the $\zeta$'s range is reduced to a point~\footnote{For example by letting $b_1\rightarrow w_2^2$.}. Solutions with no F point can be understood as generalised circular solutions discussed in section~\ref{sec31}, while those with an F point are generalised folded solutions of section~\ref{sec32}.

Finally we should consider the action of the orientifold on the above open and closed string solutions. The orientifold will act on the opens string solutions as
\begin{equation}
x_1(\sigma)\rightarrow x_1(\pi-\sigma)\,,\qquad
x_2(\sigma)\rightarrow x_2(\pi-\sigma)\,,\qquad 
x_3(\sigma)\rightarrow -x_3(\pi-\sigma)\,,\label{orientact}
\end{equation}
with the action on the closed strings as above but with $\pi$ replaced by $2\pi$. 

We have already seen in the previous sections that open string folded and circular solutions are very similar to the closed string solutions. The same turns out to be true for the more general solutions considered in this section. For consider any open string solution $x_i^o(\sigma)$ which satisfies the boundary conditions~(\ref{openbc}) and is invariant under the orientifold action~(\ref{orientact}). It is easy to see that for such a solution $x_i(0)=x_i(\pi)$, in other words that the end-points of the open string are at the same point on $S^3$. Defining a closed string solution as
\begin{equation}
x_i^c({\tilde \sigma})=x_i^o(\sigma)\,,\qquad \mbox{where } {\tilde\sigma}=2\sigma\,,\label{opentoclosed}
\end{equation}
we see immediately that $x_i^c$ satisfy the closed string periodicity conditions~(\ref{closedbc}) and are invariant under the closed string orientifold action. Similarily, consider a closed string solution $x^c_i$, which satisfies equation~(\ref{closedbc}) and is invariant under the orientifold action. It is easy to see that $x_3^c(0)=0=x_3^c(2\pi)$.
Differentiating with respect to $\sigma$, the orientifold-invariance conditions for $x_a^c$ ($a=1,2$) one also finds
that $x_i^c{}^\prime(0)=0=x_i^c{}^\prime(2\pi)$. From $x_i^c$ we can define $x_i^o$ as in equation~(\ref{opentoclosed}) which satisfies the open string boundary conditions~(\ref{openbc}) and is invariant under the open string orientifold action. 

\noindent Note that the mid-point of the open/closed string is also a B$_3$ point. A general orientifold invariant solution will then start at a B$_3$ point and go through several extrema allowed in its $\zeta$ range (as discussed above equation~(\ref{allowedextrema})), before it gets to its mid-point which is also a B$_3$ point. The solution after the mid-point is defined by requiring invariance under the orientifold action. As in the previous sections the open string solution will carry a Chan-Patton matrix $\beta$, and the full solution should be viewed as a tensor product of $x_i^o$ with $\beta$. Since $X_0=\kappa\tau\otimes\beta$ has to be invariant under the orientifold action we see that $\beta=\gamma_g\beta^T\gamma_g^{-1}$.

\noindent Open string solutions constructed from closed string solutions via equation~(\ref{opentoclosed}) will carry half the angular momenta and energy of the closed string solution. As a result we have
\be
E_{\mbox{\scriptsize o}}(J_{i,\mbox{\scriptsize o}})=\frac{1}{2}E_{\mbox{\scriptsize c}}(2J_{i,\mbox{\scriptsize c}})\,.
\label{eoec}
\ee
In other words the energy of the open string solutions can be read-off from the orientifold-invariant closed string solutions.

\noindent As an example of an orietifold invariant solution consider the folded-bent string discussed in section 3 
of~\cite{AFRT}. By relabeling $x_3\leftrightarrow x_2$ and $\sigma\rightarrow\sigma+\pi/2$ we obtain a closed string solution of the type B$_3\rightarrow$F$\rightarrow$B$_3\rightarrow$F$\rightarrow$B$_3$, which is the minimal orientifold invariant solution containing an $F$ point and, as discussed above, can be interpreted both as an open and closed string solution. The energy of the closed string solution can be found in~\cite{AFRT}, while that of the open string solution follows from equation~(\ref{eoec}).
\section{Relationship to gauge theory}\label{sec7}
In~\cite{pporf} an identification between the free open and closed string spectrum and the planar anomalous dimensions of a large R-charge sector of the gauge theory was carried out. Later, using lightcone string field theory for open strings~\cite{osft},~\footnote{For a recent paper on open-closed string field theory see~\cite{lss}.} the non-planar corrections were considered~\cite{gmp}. In the previous sections we have found that the classical open and closed spinning string solutions invariant under the orientifold action have an energy expansion of the
form~(\ref{eexp}) - in integal powers of $\lambda$, with sigma model corrections suppressed by extra powers of $1/J$.
Reproducing this expression as the exact scaling dimensions $\Delta(\lambda,J)$ of SYM operators with the same global
charges constitutes a test of the AdS/CFT duality. 

\noindent In this section we identify in more detail the map between these global charges and the operators dual to the string solutions.
On the string side the orientifold action does not act on the $AdS_5$ geometry, in other words the energy $E$ and two
spins $S_1,\,S_2$, which are Casimirs of $SO(2,4)$ are the same as in the unorientifolded theory. In particular, we may
identify the string energy with the scaling dimension of an SYM operator, as in the ${\cal N}=4$ case. The orientifold
action breaks the $SO(6)$ symmetry of $S^5$ down to $SO(4)\times SO(2)\sim SU(2)_L\times SU(2)_R\times U(1)$ where, in
our conventions, $J_Z$ is the $U(1)$ Casimir, while $J_X$ and $J_Y$ combine to give the Casimirs of the two $SU(2)$'s
\be
J_{SU(2)_R}=\frac{1}{2}(J_X+J_Y)\,, \qquad J_{SU(2)_L}=\frac{1}{2}(J_X-J_Y)\,.
\ee
The ${\cal N}=2$ $Sp(N)$ theory has a vector multiplet, whose scalar we denote by $W$, as well as a hypermultiplet in
the antisymmetric representation. This can be split into two chiral fields $Z,\,Z^\prime$. Further, there are four
hypermultiplets in the fundamental, which also possess a global $SO(8)$ symmetry, as a result of which we write them as
an $SO(8)$ vector $Q_i$. The `closed string' chiral fields $Z,\,Z^\prime,W$ carry charges $(J_X,J_Y,J_Z)$ equal to
$(1,0,0)$, $(0,1,0)$ and $(0,0,1)$ respectively, while the $Q_i$ have charges $(1/2,0,0)$.  In the planar limit, there
are two types of gauge invariant operators we may consider: those in a trace, or those sandwiched between two
fundamental fields. These correspond to closed and open strings respetively. For example the energy of the open string
solution of section~\ref{sec32}, should match with the anomalous dimension of an operator
\be
\beta_{ij}Q_i\Omega(\Omega Z)^{J_X}(\Omega Z^\prime)^{J_Y}Q_j+\cdots\,,\label{openstrop}
\ee
where $\beta$ is the $SO(8)$ `Chan-Patton' matrix, and $\cdots$ denotes all permutations of the above, which have the
same global charges and classical scaling dimension.\footnote{In keeping with standard notation, in this section
$\Omega$ denotes the invariant tensor of $Sp(N)$, rather than string orientiation reversal.} Similarily a general closed
string solution of the type discussed in section~\ref{sec6} should have an energy expansion which matches the anomalous
dimension of the operator
\be
\Tr\left((\Omega Z)^{J_X}(\Omega Z^\prime)^{J_Y}(\Omega W)^{J_Z}\right)+\cdots\,.\label{closedstrop}
\ee

Given the work on the anomalous dimensions of ${\cal N}=4$ operators~\cite{mz,gt}, it should be straightforward to find a
spin-chain model corresponding to the closed string operators - afterall these are just an orientifold-invariant
subsector of the ${\cal N}=4$ theory. The open string operators of the type~\ref{closedstrop}), should be described by a
spin-chain model which satisfies suitable boundary conditions, rather than being periodic. The string theory prediction
is that the eigenvalues of the Hamiltonian of such an open spin-chain model will be related to the eigenvalues of the
periodic spin-chain Hamiltonian via equation~(\ref{eoec}). It would be interesting to carry out this analysis explicitly.
\section{Conclusion}\label{sec8}
In this paper we have found classical open and closed string solutions in the $AdS_5\times S^5/\Zop_2$ geometry whose energy has a regular expansion in $\lambda/J^2$ with quantum sigma model corrections suppressed by extra powers of 
$1/J$~\footnote{We have concentrated on solutions with angular momenta only along $S^5/\Zop_2$, but a generalisation to solutions with non-trivial spins is straightforward since $\Zop_2$ acts trivially on $AdS_5$.}. The theory considered here has half the supersymmetries of the previously studied examples. Furthermore this is the first time that open spinning strings have been studied. We have identified the operators whose anomalous dimensions should be equal to the energy of the spinning strings; in particular, as in the BMN case~\cite{pporf}, open strings correspond to 
two-quark operators and closed strings to single-trace operators.

\noindent We have found an equivalence between classical open and closed string solutions, and their energies (see equation~(\ref{eoec})). Since the energies represent an exact expression in $\lambda/J^2$ at the planar level, this gives a prediction for an all-loop relationship between the anomalous dimensions of two-quark and single-trace operators on the gauge theory side. It would be interesting to extend the spin-chain formalism~\cite{mz,gt}, developed for the ${\cal N}=4$ 
theory, to the ${\cal N}=2$ $Sp(N)$ theory considered here.\footnote{Recently~\cite{ww} some work in this direction has been done for ${\cal N}=1,2$ gauge theories obtained by orbifolding.} In this case there should be two types of spin chains - periodic ones corresponding to single-trace operators and open ones corresponding to two-quark operators. It would be interesting to find their respective hamiltonians, and to apply the Bethe ansatz techniques to study their eigenvalues.

\noindent A generalisation of the analysis of~\cite{AS} to the case at hand is also desirable. Given the equivalence of open and closed string solutions on the string theory side, the Backlund equations should readily give an infinite set of local charges for the open strings. These should be compared with the higher local charges of the Bethe ansatz in the open spin-chain model. From the point of view of non-local symmetries too, open spin chains are perhaps more interesting than the periodic ones. For rapidly decreasing boundary conditions the non-local Yangian symmetry becomes an exact symmetry of the theory. It is then expected that given one solution, one can generate others by applying to it the non-local generators of the Yangian. Finally, all the evidence of the spinning string spin-chain duality, suggests that there should be a mapping between the two structures without reference to any particular solution. Recently, progress on this has been achieved~\cite{krucz}, and finding an exact map would be very instructive.

\vskip1cm 
\section*{Acknowledgments}
I would like to thank Jorge Russo, Matthias Staudacher and especially Gleb Arutyunov and Arkady Tseytlin for discussions, sharing their insights and comments on the manuscript. I am grateful to Kelly Stelle and the Imperial College Theoretical Physics group for their hospitality during the final stages of this project. This work was supported by FOM, the Dutch Foundation for Fundamental Research on Matter.
\appendix
\section{Stability Analysis of the Two-Spin Circular solutions}\label{sec4}
\noindent In this section we analyse the stability of the open and closed string two-spin circular solutions found in the previous section. We briefly review the analysis of~\cite{FT3} and apply it to the closed and open string solutions described in section~\ref{sec3}. In~\cite{FT3} it was found that the quadratic Lagrangian for fluctuations in $S^5$ around the closed string solution~(\ref{clsoln}) is
\be
L_2=-\p_a{\tilde X}_m\p^a{\tilde X}_m-\Lambda{\tilde X}_m{\tilde X}_m-2{\tilde\Lambda}X_m{\tilde X}_m\,.
\ee
Redefining ${\tilde X}$ as
\be
{\tilde X}(\tau,\sigma)=O_{12+34}(w\tau)O_{13+24}(k\sigma)O_{56}(\nu\tau)O_{15}(\gamma_0)\xb(\tau,\sigma)\,,
\label{xb}
\ee
it is possible to satisfy the ${\tilde\Lambda}$ constraint with $\xb_1=0$. The unconstrained Lagrangian is then
\ba
L_2&=&(\p_\tau\xb_s)^2-(\p_\sigma\xb_s)^2+4\nu\sin\gamma_0\xb_5\p_\tau\xb_6
-4w(\cos\gamma_0\xb_5\p_\tau\xb_2+\xb_3\p_\tau\xb_4)\nonumber\\
& &\qquad\qquad\qquad+4k(\cos\gamma_0\xb_5\p_\sigma\xb_3-\xb_2\p_\sigma\xb_4)\,,\label{lagfluct}
\ea
where compared to section~\ref{sec3} we have set $k\equiv n_1=n_2$, and $s=2,\dots,6$.
\subsection{Closed strings}\label{sec421}
Since the closed string is periodic in $\sigma$, we look for solutions of the form
\be
\xb_s=\sum_{n=-\infty}^\infty\sum_{i=1}^8A^{(i)}_{sn}e^{i(\omega_{n,i}\tau+n\sigma)}\,.\label{xbexpand}
\ee
Non-trivial solutions of the equations of motion following from~(\ref{lagfluct}) require the vanishing of the determinant of the characteristic matrix
\ba
\det(M_c)&=&\left(n^2-\omega_n^2\right)\left(\omega_n^4-2\omega_n^2(n^2+2(\nu^2+k^2))+n^2(n^2-4k^2)\right)
\nonumber \\
&&\qquad\qquad\left(\omega_n^4-2\omega_n^2(n^2+2(\nu^2+k^2\cos^2\gamma_0))+n^2(n^2-4k^2\cos^2\gamma_0)\right)\,,\label{detm}
\ea
This expression is the same as equation (2.27) of~\cite{FT3}. Setting $k=0$ in equation~(\ref{detm}) we recover the BMN spectrum
\be
\omega_n=\pm^\prime\nu\pm\sqrt{n^2+\nu^2}\,,
\ee
while setting $\gamma_0=\pi/2$ the frequencies are 
\ba
\omega_n^2&=&n^2+2(\nu^2+k^2)\pm 2\sqrt{(\nu^2+k^2)^2+n^2(\nu^2+2k^2)}\,,\\
\omega_n&=&\pm^\prime\nu\pm\sqrt{n^2+\nu^2}\,.
\ea
Reality of $\omega_n$ then requires $n^2\ge 4k^2$, which is not satisfied for 
$n=\pm 1\,,\dots\,,\pm (2k-1)$ and corresponds to the instability discussed in~~\cite{FT3}.

\noindent The action of the orientifold generator $g$ on the fluctutations $\xb_s$ follows from 
equations~(\ref{Iy}) and~(\ref{xb}) and is given by
\ba
\xb_s(\tau,\sigma)&\rightarrow& \xb_s(\tau,2\pi-\sigma)\,,\qquad s=1,\,2,\,5,\,6\,,\nonumber\\
\xb_s(\tau,\sigma)&\rightarrow& -\xb_s(\tau,2\pi-\sigma)\,,\qquad s=3,\,4\,.\label{xbcact}
\ea
It is easy to see that the Lagrangian~(\ref{lagfluct}) is explicitly invariant under this transformation.
The quadratic flucutations need to be invariant under this orientifold action, which requires
\ba
A^{(i)}_{s,n}=A^{(i)}_{s,-n}\,,s=1,\,2,\,5,\,6\,,\\
A^{(i)}_{s,n}=-A^{(i)}_{s,-n}\,,s=3,\,4\,.
\ea
This halves the number of independent $n\neq 0$ modes, and removes the $n=0$ modes in the
$s=3,\,4$ directions. As we saw above any potential unstable modes come as pairs $\pm n$. The orientifold projection then removes one linear combination of these modes and keeps the other. As a result the orientifold projection will not fully stabalise any unstable solutions.
\subsection{Open strings}\label{sec422}
Consider next small fluctuations around the open string solution. The fields ${\tilde X}_3$ and 
${\tilde X}_4$ will have Dirichlet boundary conditions at their end-points with the other spherical fluctuations having Neumann boundary conditions. From equation~(\ref{xb}) one can see that $\xb_3$ and $\xb_4$ will satisfy Dirichlet boundary conditions, while $\xb_1,\,\xb_2,\,\xb_5$ and $\xb_6$ will satisfy Neumann boundary conditions. 

\noindent The orientifold generator $g$ acts on ${\tilde X}_m$ as
\be
g({\tilde X}_{1,2,5,6}(\tau,\sigma))={\tilde X}_{1,2,5,6}(\tau,\pi-\sigma)\,,\qquad
g({\tilde X}_{3,4}(\tau,\sigma))=-{\tilde X}_{3,4}(\tau,\pi-\sigma)\,.
\ee
From equation~(\ref{xb}) it then follows that for $k$ even
\be
g(\xb_{1,2,5,6}(\tau,\sigma))=\xb_{1,2,5,6}(\tau,\pi-\sigma)\,,\qquad
g(\xb_{3,4}(\tau,\sigma))=-\xb_{3,4}(\tau,\pi-\sigma)\,,\label{goeven}
\ee
while for $k$ odd
\ba
g(\xb_{3,4,6}(\tau,\sigma))&=&\xb_{3,4,6}(\tau,\pi-\sigma)\,,\qquad
g(\xb_2(\tau,\sigma))=-\xb_{2}(\tau,\pi-\sigma)\,,\nonumber\\
g(\xb_1(\tau,\sigma))&=&-\cos(2\gamma_0)\xb_1(\tau,\pi-\sigma)+\sin(2\gamma_0)\xb_5(\tau,\pi-\sigma)\,,
\nonumber\\
g(\xb_5(\tau,\sigma))&=&\sin(2\gamma_0)\xb_1(\tau,\pi-\sigma)+\cos(2\gamma_0)\xb_5(\tau,\pi-\sigma)\,.
\label{goodd}
\ea
It is easy to see that the action squares to one. Note also that for $k$ odd setting $\xb_1=0$ in order to
satisfy the ${\tilde\Lambda}$ constraint is not a $g$ invariant process.
However, it was argued below equation~(\ref{gopen}) that the classial solution with $k$ odd is not $g$ invariant
and so we are not interested in fluctuations around it.

\noindent For $k$ even we may set $\xb_1=0$, and look for solutions to the Lagrangian~(\ref{lagfluct}) of the form
\ba
\xb_s&=&\sum_{n=0}^\infty\sum_{i=1}^8A^{(i)}_{sn}e^{i\omega_{n,i}\tau}\cos(n\sigma)\,,\qquad s=2,\,5,\,6 \\
\xb_s&=&\sum_{n=1}^\infty\sum_{i=1}^8A^{(i)}_{sn}e^{i\omega_{n,i}\tau}\sin(n\sigma)\,,\qquad s=3,\,4\,.
\ea
The determinant of the characteristic matrix is the same as in the closed string case and so the frequencies will be the same as in the closed string case. In particular, in the un-orientifolded theory, the stability analysis is identical to the closed string case.

\noindent Taking the orientifold of the above fluctuations is also straightforward
\ba
g(\xb_s)&=&\sum_{n=0}^\infty\sum_{i=1}^8(-1)^nA^{(i)}_{sn}e^{i\omega_{n,i}\tau}\cos(n\sigma)\,,\qquad s=2,\,5,\,6 \\
g(\xb_s)&=&\sum_{n=1}^\infty\sum_{i=1}^8(-1)^nA^{(i)}_{sn}e^{i\omega_{n,i}\tau}\sin(n\sigma)\,,\qquad s=3,\,4\,,
\ea
which needs to be combined with the action of $g$ on the $SO(8)$ Chan-Patton factors ${\bar\beta}$. As a result $n$-even fluctuations will have Chan-Patton factors which satisfy ${\bar\beta}=\gamma_g{\bar\beta}^T\gamma_g^{-1}$, while the
$n$-odd fluctuations will have Chan-Patton factors which satisfy ${\bar\beta}=-\gamma_g{\bar\beta}^T\gamma_g^{-1}$.
Since ${\bar\beta}$ are $8\times 8$ matrices, there will always be some $g$-invariant states for any value of $n$.
Hence, the open string solutions with $k$ even, will be stable in the same range of parameters as the closed string solutions.

\end{document}